\documentclass[12pt]{iopart}
\usepackage{graphicx}

\begin{document}

\title{Exotic phenomena in doped quantum magnets}

\author{D.~Poilblanc$^1$, M.~Mambrini$^1$, A.~L\"auchli$^2$, F.~Mila$^3$}

\address{$^1$
Laboratoire de Physique Th\'eorique,
CNRS \& Universit\'e de Toulouse,
118 rte de Narbonne, 31062 Toulouse, France}

\address{$^2$
  Institut Romand de Recherche Num\'erique en Physique des Mat\'eriaux (IRRMA),
  PPH-Ecublens, CH-1015 Lausanne, Switzerland}

\address{$^3$
Institute of Theoretical Physics,
Ecole Polytechnique F\'ed\'erale de Lausanne,
BSP 720, CH-1015 Lausanne, Switzerland}

\ead{didier.poilblanc@irsamc.ups-tlse.fr}

\begin{abstract}
We investigate the properties of the two-dimensional
frustrated quantum antiferromagnet on the square lattice,
especially at infinitesimal doping.
We find that next nearest neighbor (N.N.) $J_2$ and 
next-next N.N. $J_3$ interactions together destroy the
antiferromagnetic long range order  
and stabilize a quantum disordered valence bond 
crystalline {\it plaquette}
phase. A static vacancy or a dynamic hole doped into
this phase liberates a spinon.
From the profile of the spinon wavefunction around the (static) vacancy 
we identify an
intermediate behavior between complete deconfinement (behavior seen
in the kagome lattice) and
strong confinement (behavior seen in the checkerboard lattice) 
with the emergence
of two length scales, a spinon confinement length {\it larger}
than the magnetic correlation length. When a finite hole hopping
is introduced, this behavior
translates into an extended (mobile) spinon-holon boundstate with a very 
small quasiparticle weight. These features provide clear evidence
for a nearby "deconfined critical point" in a doped microscopic
model. Finally, we give arguments in favor of superconducting 
properties of the doped plaquette phase. 
\end{abstract}


\submitto{\JPCM}

\maketitle
\section{Introduction: the plaquette valence bond crystal}

\begin{figure}
\centerline{\includegraphics*[width=0.9\linewidth]{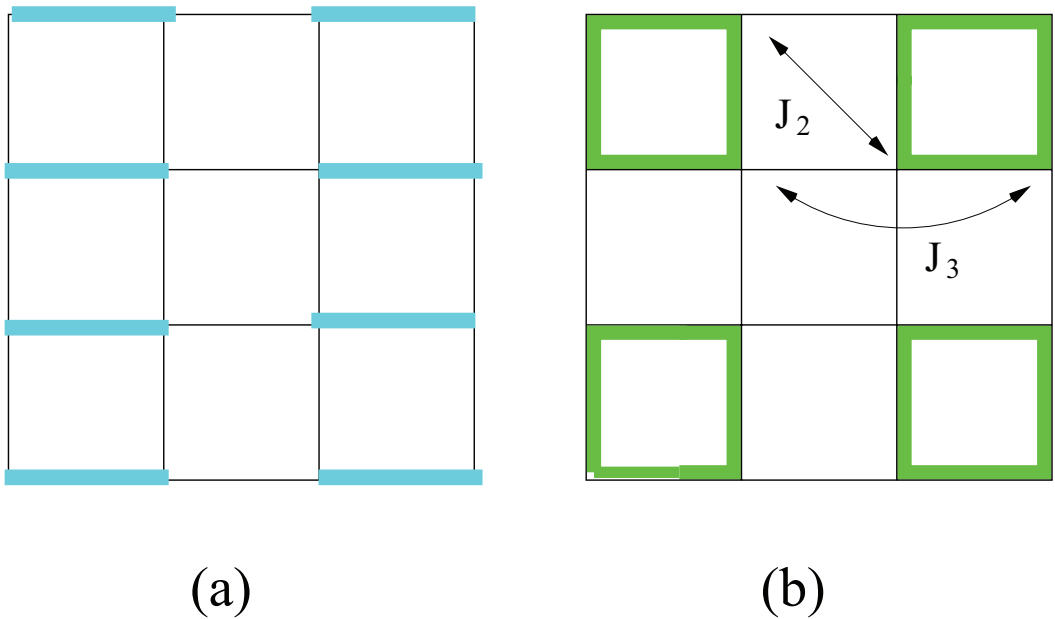}}
  \caption{\label{fig:VBC}
(Color on-line) Exemples of simple VBC states,
(a) columnar phase and (b) plaquette phase, where 2 (4) spins 
are paired up in dimer- (plaquette-) singlets. 
In both cases the ground state is 4-fold degenerate.
The $J_2$ and $J_3$ couplings 
(depicted on figure (b)) are shown to stabilize the plaquette phase.
}
\end{figure}

Magnetic frustration is believed to be the major tool
to drive a two-dimensional (2D) quantum antiferromagnet
(AF) into exotic quantum disordered SU(2)-symmetric phases such as
the spin liquid (SL) state characterized by the absence of ordering
of any kind, and possibly observed in the 2D kagome
lattice~\cite{Kagome}. The Valence Bond Crystal
(VBC) which, in contrast, breaks
lattice symmetry (see Fig.~\ref{fig:VBC} for pictures of such states), 
seems to be a serious alternative in some other
frustrated quantum magnets as suggested by robust field
theoretical arguments~\cite{VBS}, early numerical computations of
frustrated quantum AF on the square lattice with diagonal
$J_2$ bonds~\cite{j1j2} and
in the 2D checkerboard lattice~\cite{checkerboard} (with diagonal
bonds only on half of the plaquettes). 

The "deconfined critical point"
(DCP)  was recently proposed to describe
a new class of quantum criticality characterizing
the AF to VBC transition~\cite{Senthil}.
Here we argue that the two-dimensional (2D) spin-1/2 AF 
$J_1$-$J_2$-$J_3$ Heisenberg model~\cite{Ferrer} on the square lattice
defined by
\begin{equation}
   H= \sum_{\langle ij\rangle}J_{ij}{\bf S}_i \cdot {\bf S}_j
\label{eq:H}
\end{equation}
where the $J_{ij}$ exchange parameters
are limited to first ($J_1$), second ($J_2$)
and third ($J_3$) N.N. AF couplings is a strong candidate for 
exhibiting such a transition. Recent investigations of this 
model along the pure $J_3$ axis~\cite{Leung,CapriottiSachdev} or 
for both $J_2$ and $J_3$ finite~\cite{Mambrini_j1j2j3} provide
an accumulation of evidence in favor of
the existence of a VBC phase with strong plaquette correlations
similar to the cartoon of Fig.~\ref{fig:VBC}(b). 

\begin{figure}
\centerline{\includegraphics*[angle=0,width=0.7\linewidth]{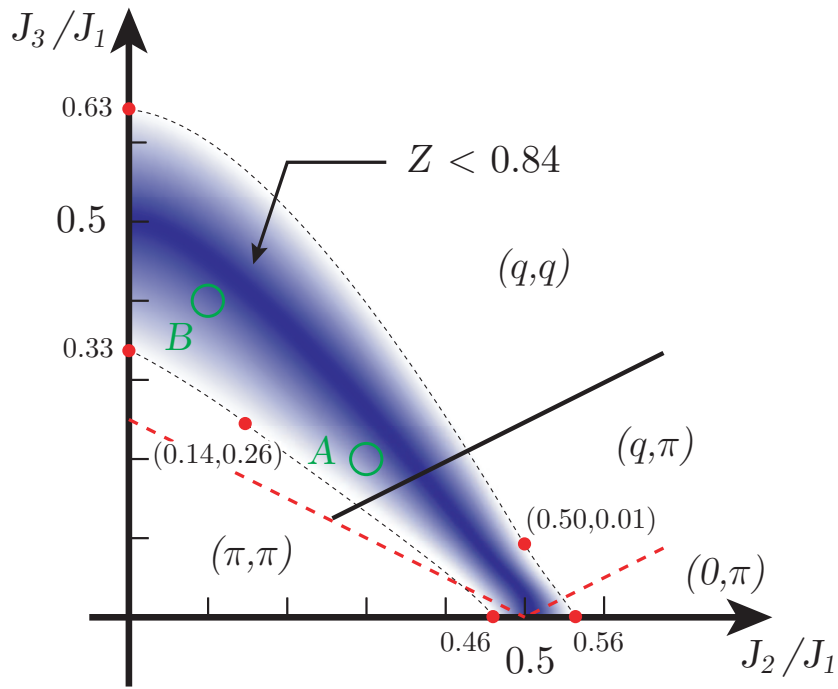}}
  \caption{\label{fig:phase_diag}
(Color on-line) Classical phase diagram for the $J_1$-$J_2$-$J_3$
model. Second order (discontinuous) transitions are indicated by
dashed (solid) lines (see Ref.~\protect\cite{Ferrer} for more details).
The shaded (blue online) region shows the approximate location of the
minimum of the impurity spectral weight $Z$ ($0.79<Z<0.84$) 
in the quantum version.
}
\end{figure}

\begin{figure}
\centerline{\includegraphics*[angle=0,width=0.7\linewidth]{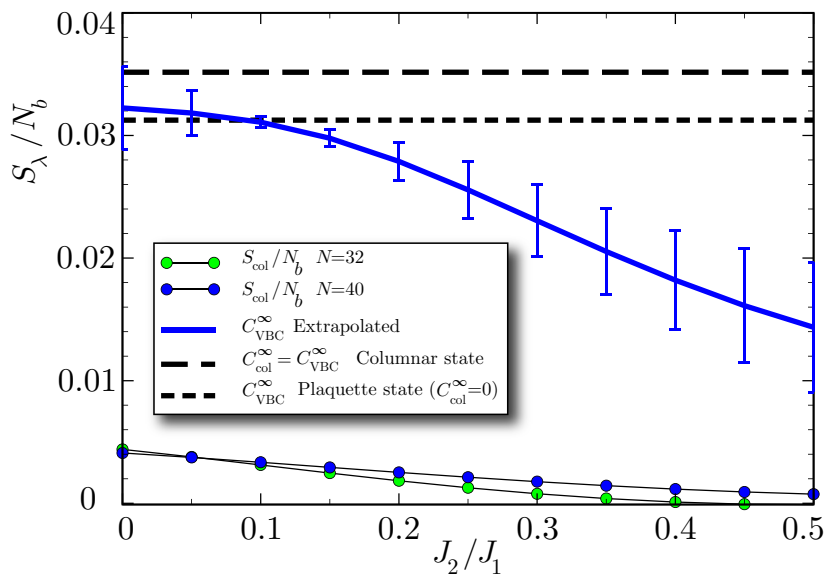}}
\caption{\label{fig:struct_factors} (Color on-line).
VBC (extrapolation) and columnar (for $32$ and $40$ sites clusters) 
structure factors 
along a simple cut $J_2+J_3=1/2$ in parameter space (see text).
The dashed lines correspond to the value of both structure factors
expected in pure columnar dimer and plaquette states. From 
Ref.\protect\cite{Mambrini_j1j2j3}.
  }
\end{figure}

Extensive exact diagonalisations of the model within its full Hilbert 
space and within a restricted Hilbert space of N.N. (SU(2)) dimer
configurations reported in Ref.~\cite{Mambrini_j1j2j3} strongly support
the existence of a plaquette VBC phase along the $J_2+J_3=1/2$ 
line of parameter space (see Fig.~\ref{fig:phase_diag}).
We shall provide here a selection of these data to illustrate
this point. Dimer-dimer correlations (not shown) 
reveal a strong signal of 
VBC order of some kind. Summing up these spatial correlations with appropriate 
phase factors provides quantitative estimates of both the 
''generic'' VBC and the ''specific'' columnar VBC structure factors 
(normalized to give the squares of the corresponding order parameters). 
Such estimates are displayed in  Fig.~\ref{fig:struct_factors};
the upper line is obtained by an extrapolation to infinite size
of the dimer structure factor constructed by summing up the correlations of the
horizontal dimers (only) with an alternating sign along columns (in order to 
provide a signal for all types of VBC phase). Clearly, a
large VBC order parameter is seen with a value close to
the one of a pure plaquette phase (dotted line), 
at leat close to the pure $J_3$ axis.
The data sets at the bottom are obtained by summing up the 
dimer correlations with opposite sign for vertical and horizontal dimers
(also omitting the short distance contributions) to filter the signal
of columnar order only. The small values of the latter data 
strongly suggest that the VBC order is a plaquette order 
rather than a columnar order. Note, however, that the VBC signal
weakens significantly when approaching the pure $J_2$ axis. 

\section{Doping the plaquette VBC}

We shall now test the properties of the plaquette VBC by removing an electron 
at a given site (e.g. by chemical substitution
with an inert atom) or, as in ARPES
experiments, in a Bloch state of given
momentum. This process naturally liberates a spinon,
i.e. a S=1/2 polarisation in the vicinity of the empty site (holon).
A new length scale $\xi_{\rm conf}$ corresponding to the average distance
between vacancy and spinon emerges
naturally in a VBC phase and is to be identified
with the correlation length over which VBC order sets in.
Interestingly, it has been predicted that,
in the vicinity of the DCP,
confinement occurs on a much larger length scale
$\xi_{\rm conf}$ than the spin-spin correlation 
length $\xi_{\rm AF}$~\cite{Senthil}.

\begin{figure}
\centerline{\includegraphics*[angle=0,width=0.7\linewidth]{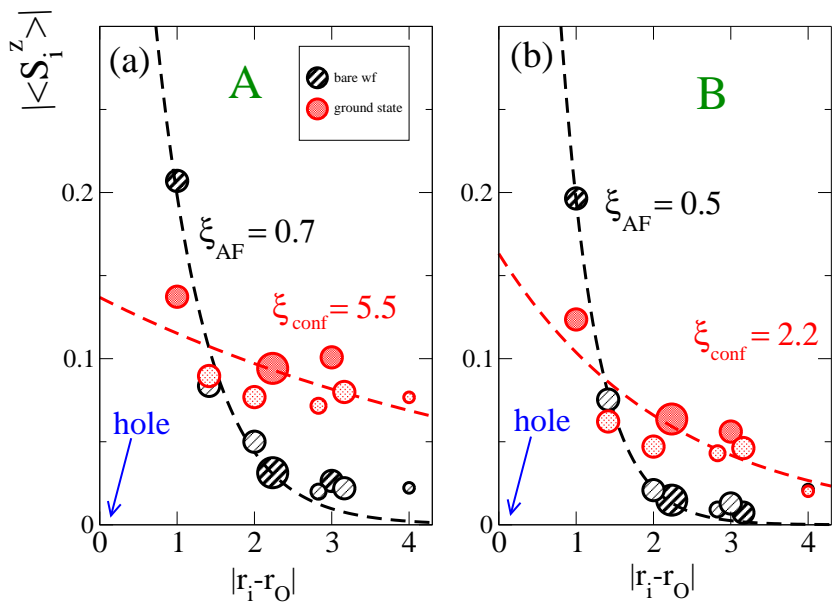}}
\caption{\label{fig:siz2} (Color on-line) Modulus of the
spin-spin correlation (black)
and of the spin polarization
in the vicinity of the vacancy (red) (summed up on equivalent sites) 
versus distance 
for both $J_2/J_1=0.3$ and $J_3/J_1=0.2$ (a) and $J_2/J_1=0.1$ and
$J_3/J_1=0.4$ (b) corresponding to points A
and B
in the phase diagram of
Fig.~\protect\ref{fig:phase_diag}. Fits using exponential forms
are shown in dashed lines. The areas of the dots are proportional
to the number of equivalent sites (entering in
the fits). Dark and light symbols correspond to positive and negative values
respectively. From ref.~\protect\cite{Poilblanc_j1j2j3}.
  }
\end{figure}

In Fig.~\ref{fig:siz2} we compare the decay of the spin-spin correlation 
with distance to the one of the spinon ''cloud'' away from the vacancy
which both can be fitted usig exponential forms, hence enabling to 
extract the corresponding length scales.
The obtained very short magnetic correlation length
$\xi_{\rm AF}$, below one lattice spacing, is to be contrasted
with the strikingly large {\it confinement} length $\xi_{\rm conf}$
typically ranging from 2 to 6 lattice spacings.
This is to be compared with two other behaviors observed in the kagome and 
in the checkerboard lattices~\cite{Poilblanc_j1j2j3} respectively: 
in the checkerboard lattice 
(whichalso exhibits a plaquette phase but a larger spin 
gap~\cite{checkerboard}) very 
short-ranged spin-spin correlations are seen while the spinon remains 
almost entirely confined on the
N.N. sites of the vacancy. In contrast, on the kagome lattice, the 
spin-1/2 delocalizes on the whole lattice, a clear signature of 
deconfinement which is consistent with a SL ground state.

\begin{figure}
\centerline{\includegraphics*[angle=0,width=0.65\linewidth]{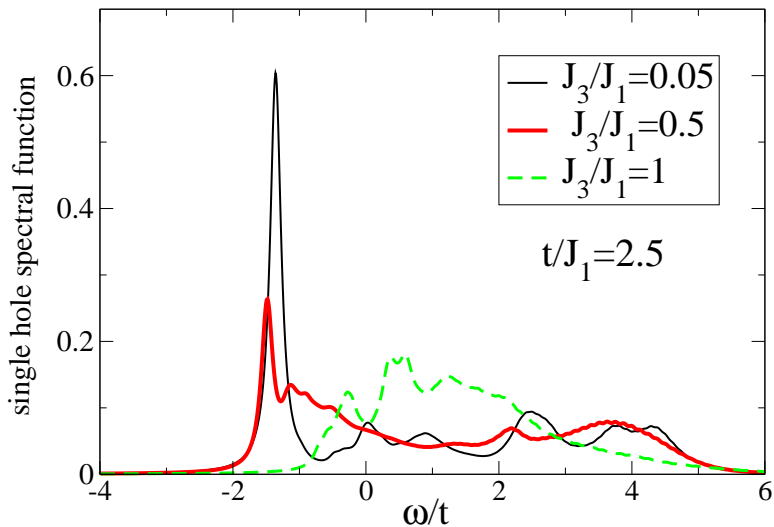}}
\caption{\label{fig:spectral2} (Color on-line).
Single hole spectral function (at momentum $(\pi,0)$) for
various values of the magnetic frustration and a realistic hole hopping of
$t/J_1=2.5$. Note that for $J_3/J_1=0.5$ the main low energy peak contains several poles in contrast to the $J_3=0.05$ case where a single pole is present. 
  }
\end{figure}

The quasiparticle weight Z can be deduced from the computation of the 
single hole Green function. Fig.~\ref{fig:phase_diag} 
shows that it is clearly reduced in the 
vicinity of the J2+J3 = J1/2 line where evidence for a plaquette VBC phase
was found. This behavior is in agreement with the above real-space picture 
which naturally implies that the extended spinon wavefunction of 
size $\xi_{\rm conf}$ has a reduced overlap $Z$ with the bare wavefunction 
(of extension $\xi_{\rm AF}$). When the vacancy is given some kinetic energy 
(through a N.N. hopping as in the well-known t-J model), the $Z$ 
factor of the hole drops further to very small values~\cite{Poilblanc_j1j2j3} 
showing that the hole motion strongly suppresses the effect of the remaining 
spinon-holon VBC string potential~\cite{Laughlin}. Typical spectral functions 
along the pure J3 line are shown in Fig.~\ref{fig:spectral2}. 
Note that a {\it complete} 
deconfinement was shown for a mobile hole on the kagome lattice~\cite{PRL} 
as evidenced by a fully incoherent spectral weight at low energies.

Lastly, we would liketo comment on the possibility of superconducting 
pairing upon doping VBC's. In fact, pairing driven by a kinetic energy 
gain of the Cooper pair was discovered in the robust plaquette phase 
of the checkerboard lattice~\cite{pairing_VBC}. On general grounds,
one would also 
expect hole pairing in the doped J$_1$-J$_2$-J$_3$ 
quantum antiferromagnet due to the long 
distance confining string of the VBC. However, short range effects 
(as in the checkerboard lattice) could boost the weak pairing interaction.

\section{Conclusions}

To conclude, the confinement of a spinon liberated by introducing a vacant 
site or a mobile hole has been studied in the J$_1$-J$_2$-J$_3$ model where 
a plaquette VBC phase has been identified in some extended region of the phase 
diagram. In this region of large frustration, we have identified a new 
length scale
related to the confinement of the spinon. Its large value compared to the 
spin-spin correlation length supports the field-theoretic
``deconfined critical point scenario''~\cite{Senthil} for the N\'eel-VBC 
transition.

\end{document}